# Frozen steady states in active systems

Volker Schaller[1], Christoph Weber[2], Benjamin Hammerich[1], Erwin Frey[2] and Andreas R. Bausch[1]

[1]*Lehrstuhl für Biophysik-E27, Technische Universität München, Garching, Germany*
[2]*Arnold Sommerfeld Center for Theoretical Physics and Center for NanoScience, Department of Physics, Ludwig-Maximilians-Universität München, Munich, Germany*

**02.09.2011**

**Even simple active systems can show a plethora of intriguing phenomena and often we find complexity were we would have expected simplicity. One striking example is the occurrence of a quiescent or absorbing state with frozen fluctuations that at first sight seems to be impossible for active matter driven by the incessant input of energy. While such states were reported for externally driven systems through macroscopic shear or agitation, the investigation of frozen active states in inherently active systems like cytoskeletal suspensions or active gels is still at large. Using high density motility assay experiments, we demonstrate that frozen steady states can arise in active systems if active transport is coupled to growth processes. The experiments are complemented by agent-based simulations which identify the coupling between self-organization, growth and mechanical properties to be responsible for the pattern formation process.**

How does a dynamic system approach its steady state and what is the nature of it? For a thermodynamic system Onsager's regression hypothesis (1) asserts that the relaxation is governed by the same laws as the fluctuations in thermal equilibrium, and Boltzmann-Gibbs theory defines the nature of all steady states. For active systems like vibrated granules (2, 3), animal swarms (4-6), microorganisms (7-9) and cytoskeletal systems (10-13) there are no such general laws. Instead one finds a plethora of spatio-temporal patterns including swarming (5-8), density inhomogeneities (11, 13), and swirling patterns (9, 12). A common feature of these patterns is that they are all fluid-like (14-18). One might ask whether there is something like the analog of a solid-like or frozen state in active systems. At first sight this seems to be impossible for active matter driven by the incessant input of energy which is expected to cause highly dynamic states through a perpetual balance between assembly and disassembly processes (13). Yet, there are examples where the constituents of active systems self-organize into non-equilibrium steady states with slowed dynamics or regular spatial patterns. Activity has been found to give rise to a diverging viscosity at the isotropic-nematic transition of active gels (19-21), and glassy behavior has recently been observed for confluent cell layers (22). Systems that are externally driven, like agitated granular media, exhibit a plethora of regular static patterns (23-25) and rotating assemblies (26). While in all these non-equilibrium systems degrees of freedom are slowed down or frozen out, the very existence of a pattern or the anomalous rheological properties is only made possible by the activity in the system. This has to be contrasted with non-equilibrium systems where the system ends up in a set of configurations from which it cannot escape. Once such an absorbing state is reached the dynamics ceases and the ensuing pattern is frozen in (27, 28). Particularly illustrative examples are growing bacterial colonies showing natural selection (29) or periodically sheared colloidal suspensions which self-organize into an absorbing state where particles no longer collide but simply retrace their trajectories under the external drive (30).

Here we ask for the existence of steady states which carry the hallmarks of both an active system and a frozen absorbing state, i.e. where fluctuations are successively eliminated during a coarsening process while keeping the system in an active state. In driven systems frozen steady states can arise when particles are not only actively transported, but also integrated into new self-assembled higher-order structures where single particle fluctuations are arrested. A frozen active state combines the rapid self-organization characteristics of driven systems with the robustness, stability and reproducibility of growth and self-assembly processes. Thus the combination of self-organization and growth that characterizes frozen active states may turn out to be important for many systems in biology and materials science. However, the complexity of most experimental systems defies the thorough identification and analysis of the intricate balance of the underlying principles that lead to a frozen active state.

To shed light on the mechanisms that govern the emergence of such states, we introduce a reconstituted system, where highly concentrated actin filaments are actively transported and crosslinked in the two-dimensional geometry of a motility assay. We show that the interplay of only three components, actin filaments, HMM motor proteins and fascin crosslinkers is sufficient for the emergence of a frozen active steady state that consist of highly symmetric structures, rings and elongated fibres that are actively assembled and propelled by the motor proteins. The reconstituted approach complemented by agent-based simulations allows us to correlate the formation of a frozen steady state with the mechanical properties of the emergent structures and enables to identify the crosslinker mediated growth processes as driving mechanism for the emergence of frozen active steady states.

## Results and Discussion

The system consists of actin filaments and fluorescently labeled reporter filaments that are propelled by non-processive motor proteins (HMM), immobilized on a glass cover slide (11, 12). Without crosslinkers and above a critical filament density the filaments self-organize to form coherently moving structures such as clusters and density waves (12). To investigate how the dynamic structure formation is affected by a defined growth process into higher order structures, the crosslinker protein fascin is added and its concentration systematically varied. In this context fascin is ideally suited as it assembles polar filament structures and thus does not obstruct the material transport by the molecular motors.

The addition of trace amounts of fascin does not suffice to change the pattern formation mechanisms in the system. Even at high filament densities the individual filaments display directional fluctuations that stem from the persistent random walk of single filaments in the motility assay. These fluctuations are responsible for the formation of density inhomogeneities that accumulate to form coherently moving structures such as clusters and density waves (fig. 1*A*). Low fascin concentrations dampen these fluctuations by crosslinking events but are not sufficient to completely inhibit them. As a consequence structures like density waves are less pronounced (fig. 1*A*).

This drastically changes, if the added fascin concentration exceeds a critical concentration: Now the crosslinker leads to the emergence of rotating polar actin-fascin structures, which are effectively planar. The rotating speed of the rings corresponds to the single filament speeds and is about 3 μm/sec. While structure formation in the high density motility assay relies on the balance between assembly and disassembly processes, the addition of crosslinker molecules promotes defined assembly processes and at the same time inhibits the disassembly of the emergent structures. This finally leads to a quiescent steady state where all filaments are firmly incorporated in constantly rotating rings (fig.1*B* and supplemental video S1); fluctuations on the single filament level are completely arrested. The absence of disassembly pathways renders the structure formation mechanism reminiscent to a coarsening process into an absorbing state (28) and the conditions during the pattern formation are directly reflected and 'frozen-in' in the steady state.

In the frozen steady state constantly rotating rings are homogeneously distributed throughout the motility assay (fig.1*B*). No preferred direction of rotation is observable. Rings occur in two distinct conformations, which are equally abundant: open and closed (fig.1*B*). Closed rings consist of self-contained and constantly rotating actin-fascin fibres. Open rings also move on a stable circular trajectory with uniform curvature radius, as can be seen in a time overlay (fig. 1*B*). A characteristic distribution of ring curvature radii $p(r)$ in the steady state is observed (fig. 2*A*). The radii are broadly distributed with a decay towards large radii *r*. $p(r)$ shows a pronounced maximum at 9.5 μm which is of the order of the persistence length of individual filaments. Towards small curvature radii, the distribution is characterized by a cut-off radius of $r_c \approx 5$ μm, below which no rings are found. The decay of the distribution for large radii is of double exponential shape (fig. 2*A*). The double exponential nature of the curvature radii distribution is highly robust upon parameter variation and is conserved throughout variations of the fascin and actin concentrations (supplemental figure S1), making this to a generic feature of the system.

The ring curvature distribution reflects the balance between aggregation processes and active transport (fig. 2*B,C* and supplemental video S2). Polar actin-fascin strings nucleate from individual actin filaments crosslinked by fascin. While being transported, the actin strings grow larger by two competing mechanisms, either by taking up individual filaments alongside the bundle, or by merging with adjacent structures, leading predominantly to an increased length. These growth processes in turn affect the mechanical properties of the structures. A growth in length results in long and thin strings with relatively low persistence lengths. The continuous growth alongside the strings, yields shorter and thicker structures with a higher persistence length. It is the resulting increased stiffness of the individual strings which determines the susceptibility towards directional changes in the motility assay. Directional changes that alter a given curvature happen more seldom and have less effect for thicker and thus stiffer strings. This can be seen in the trajectory of individual fibres that is composed of circular segments (supplemental figure S2).

The interplay between active transport and the mechanical properties of the emergent structures determines the actual ring formation processes, and leads to two generic ring formation mechanisms: Once strings have grown enough in length, they may close on themselves (fig. 2*B*). Alternatively, lateral growth may increase the string's stiffness to such an extent that their current curvature freezes in without a ring closure and open rings result (fig. 2*C*). While the ring closure leads to predominantly small ring diameters, the frozen-in curvature process naturally yields rings with larger curvature radii. This is reflected in cumulative curvature radii distributions $P(r)$, which are different for closed and open rings. Both are exponentially distributed, yet their decay lengths differ by a factor of 3 (fig. 2*A*).

Thus the two growth processes competing for material in the active system not only lead to two ring populations – open and closed – but also directly affect the properties of the steady state by determining the size of the structures. Competition between continuous growth and merging processes of actively transported strings can readily be described using agent-based simulations (fig. 3*A*, for details see materials and methods). These two competing aggregation processes already suffice to retrieve the coexistence of open and closed rings (fig. 3*B* and supplemental video S3). Omission of one of these processes leads to the formation of either closed or open rings only. The cumulative radii distributions $P(r)$ for open and closed rings decay exponentially in accordance with experimental observations (fig. 3*C*). Moreover, the simulations allow for a backtracking of the steady state properties to the inherent noise in the active system which determines the stochasticity of each string's trajectory: Lowering the amplitude or increasing the rate of curvature changes leads to an increase in the fraction of open to closed rings, $\Gamma$ (fig. 3*D*).

This can directly be tested in the experiment as the noise level in the system can be addressed by varying the motor density on the cover slip $\sigma_m$. A decreased number of motor proteins on the surface leads to more rugged trajectories of the strings and hence to smaller curvature radii (fig. 4 *A,B*). In accordance with simulations a gradual decrease of the motor density shifts the fraction $\Gamma$ from open to closed rings: Small motor densities favour the emergence of closed rings with a $\Gamma$ of 0.5, while high motor densities lead to predominately open rings with $\Gamma = 2$ (fig. 4*C*).

Importantly, upon changing the motor density, the conformational statistics of closed rings remains invariant, while the distribution of open rings is shifted to smaller radii with smaller decay lengths (fig. 4*D*). This is attributed to the underlying coupling of growth and active transport through the mechanical properties of the emergent structures and the freezing mechanism that reflects the conditions during the assembly process. At low motor densities the less persistent movement (fig. 4*A*) implies smaller frozen-in radii of the open rings. The ring closure process, on the other hand, is triggered by stochastic changes in a string's trajectory, yet the radius is determined by the stiffness of each individual structure. As a consequence, closed rings get only more abundant but the ensuing distribution of curvature radii remains independent of the motor density $\sigma_m$.

Frozen-in structures directly document the conditions during the coarsening process, and thus are expected to strongly depend on the nucleation and growth mechanisms. In the present system these are governed by the filament and the crosslinker concentration. Increasing these concentrations not only leads to more nucleation seeds and thus actin-fascin strings but also to a more rapid growth process yielding stiffer strings. Above a critical material density at the surface, the strings are unable to pursue curved trajectories anymore and they are forced to align. Successively, they get increasingly interconnected and crosslinked to form frozen elongated and straight polar actin-fascin streaks with a thickness of up to 50 μm and a length in the order of centimetres (fig. 5 and supplemental video S4). The emergence of elongated structures following straight trajectories has severe consequences for the further growth process in the system. Compared to the rings the straight-moving streaks cover a large area. This not only leads to a more effective growth process and larger structures but also gives way to coarsening and subsequent merging processes. While neighbouring fibres initially tend to move in opposite directions, they gradually synchronize their directions of motions by merging (fig. 5*C*). This is accompanied by a coarsening process, in the course of which fibres of the same polarity gradually merge to larger ones (fig. 5*B*) which finally leads to large-scale symmetry breaking and a preferred direction of motion develops. After a time period of 10 min extended fibre structures uniformly move in the same direction, imposing a polarity in the entire flow chamber on the centimetre scale.

The phase boundary determining the transition from ring formation to fibres depends on both the actin and the fascin concentration (supplemental figure S3). Consequently, the critical material density is not related to the critical filament densities in the ordinary high density motility assay without crosslinkers.

## Conclusion

The minimal system presented here allows a systematic parameter control through the variation of motor, crosslinker and actin concentrations. This is mandatory for the robust and reproducible assembly of frozen steady states with structures of defined size and morphology. It opens the door for various applications with the nanopatterning of surfaces being a prime example, as suggested in motility assay experiments with microtubules (31, 32).

At the same time the high-density motility assay can serve as a versatile model system to explore the full breath of non-equilibrium steady states in active systems. In previous investigations it was shown that the interplay between assembly and disassembly of driven filament leads to dynamic patterns like swirls, clusters and density waves (12). These non-equilibrium steady states are characterized by the perpetual built-up and destruction of structures driven by the incessant input of energy at the scale of an individual fibre. Upon adding a single new ingredient, namely passive crosslinking molecules, we have found here that the nature of the non-equilibrium steady state changes fundamentally; the presence of crosslinking molecules facilitates permanent filament aggregation and thereby switches off the disassembly pathway. As a consequence, the system's dynamics drives the filament assembly into an absorbing state where the structure arrest while the filaments still move. The coarsening process towards this absorbing state combines active driving with filament aggregation. Once reached, this state is stable and independent of the activity of the system, yet it directly maps the assembly pathway. This 'structural memory' relies on the intricate mechanical coupling between active transport and aggregation processes. This coupling and the ensuing aggregation mechanisms fully determine the statistical properties of the absorbing state.

From the experimental point of view, it remains a great challenge to unambiguously pinpoint the nature of the non-equilibrium phase transition. There exist only very few experimental systems that show clear evidence for an absorbing phase transition – with turbulent liquid crystals being probably the first example (33). The main difficulty lies in the quantitative measurement of the transition over long time and length scales in the absence of interfering long-range interactions or boundary effects (33). In this context the system introduced here could provide a versatile tool towards the investigation of absorbing transitions. Consisting of only a few purified components, it allows for a precise investigation of the statistic properties of the phase transition. Future investigations might also want to explore analogies and differences between these active systems and externally driven systems showing shear-induced gelation (34) or aggregation (35). In general, the further investigation of active systems complemented by an assembly process that in turn affects the mechanical properties may set a new paradigm of frozen active states and seems to be an attractive route of explanation for many examples from cytoskeletal to colloidal systems.

## Materials and Methods

**Protein Preparation.** G-actin was obtained from rabbit skeletal muscle following a standardized protocol (36, 37). Actin polymerization was initiated by adding one-tenth of the sample volume of a tenfold concentrated F-buffer (20 mM Tris, 20 mM $MgCl_2$, 2 mM $CaCl_2$, 2 mM DTT and 1 M KCl). For fluorescence microscopy fluorescently labelled reporter filaments are used at a ratio of labelled to unlabeled filaments ranging from 1:2 to 1:50. They were stabilized with Alexa-Fluor-488-phalloidin (Invitrogen) at a ratio of 1:2. Unlabeled filaments are stabilized with phalloidin (Sigma) likewise at a ratio of 1:2. Once polymerized, actin was used within 2 days. HMM is prepared from myosin II obtained from rabbit skeletal muscle following a standardized protocol (38). Recombinant human fascin is purified from *E.coli* BL21-CodonPlus-RP and stored at -80°C in 2mM Tris/HCl (pH~7.4), 150 mM KCl at 64 µM, following Ref.(39).

**Sample Preparation.** Flow chambers were prepared with pre-cleaned microscope slides (Carl Roth, Germany) and coverslips (Carl Roth, Germany, 20 x 20 mm, No. 1). Coverslips were coated with a 0.1% nitrocellulose solution diluted from a 2 % solution (Electron Microscopy Sciences, Hatfiled, PA) in amyl acetate (Carl Roth, Germany). The coverslips were fixed to the microscope slides using parafilm, yielding an overall chamber volume of ~30µl. Prior to each experiment a 30µl actin dilution (1 – 25 µM monomeric actin) was prepared by gently mixing labelled and unlabelled actin filaments with Assay Buffer (25 mM Imidazolhydrochlorid pH 7.4; 25 mM KCl; 4 mM $MgCl_2$; 1 mM EGTA; 1 mM DTT). The flow chamber was incubated with HMM diluted in Assay Buffer. Prior to the insertion of the respective actin dilutions, surfaces are passified with a BSA solution (1mg/ml BSA (Sigma) dissolved in Assay Buffer). To start the experiment, 2mM of ATP together with the respective fascin concentration dissolved in Assay Buffer (50µM) is flushed into the flow chamber. Oxidation of the fluorophore was prevented by adding a standard antioxidant buffer supplement GOC (2 mg Glucose-Oxidase, Sigma; 0.5 mg Catalase, Fluka). To verify that the observed patterns are not affected by the use of GOC, control assays without the antioxidant supplement were performed. After addition of the Motility Buffer, flow chambers were sealed with vacuum grease (Bayer Silicones).

**Actin and HMM concentrations.** The numbers for the actin concentration denote the monomeric actin concentration inserted into the flow chamber. This concentration is slightly lowered by up to 10% by the subsequent rinse with the ATP-fascin dilution. The HMM concentrations $\sigma_m$ given in the manuscript denote the concentration with which the flow chamber was incubated. The relation between $\sigma_m$ and the motor density at the surface $\Psi$ was determined to increase linearly in a wide concentration range with a ratio of: $\sigma_m/\Psi = 0.04$ nM µm$^2$ (40). Above a HMM concentration of $\sigma_m =$ 200 nM, the motor density at the surface saturates at a value of $\Psi \sim 6000$ µm$^{-2}$ (41). If not indicated otherwise all experiments were performed at a HMM concentration of $\sigma_m = 90$ nM.

**Image Acquisition and Preprocessing.** All data are acquired on a Leica DMI 2000 inverted microscope with a ×63 oil objective (numerical aperture: 1.4) or a × 40 oil objective (numerical aperture: 1.35). Images (resolution: 1344 x 1024 pixels) were captured with a charge-coupled device camera (C4880-80, Hamamatsu) attached via a 0.35 or 1.0 camera mount. Image acquisition and storage are carried out with the image processing software 'OpenBox' (42).

**Ring Curvature Statistics.** For each investigated parameter set the ring curvatures were recorded in up to 20 spots in different samples. To extract the ring curvatures for each spot 200 frames with a frame rate of 8.5 frames/sec were recorded. After background subtraction (rolling ball radius of 50 pixels) we performed an average intensity projection of all 200 frames and converted the resulting picture into a binary image with a similar cutoff for all samples. All these procedures were carried out with ImageJ. The subsequent ring recognition and the ellipsoid fit to the recognized structures were carried out with a custom written MATLAB routine. All signatures which are not of approximately circular shape (ratio of semi-minor to semi-major axis <0.7) are rejected.

**Simulations.** The competition between continuous growth and merging processes is described using an agent-based simulation. Here the experimentally observed actin-fascin strings are modelled as elongated, polar strings that move with a velocity *v* on meandering trails (fig. 3a). The heads of these strings pursue circular trajectories with stochastically varying curvatures and the tail strictly following the head's trajectory. Further, the strings are subjected to aggregation processes which result in a continuous string thickening and merging processes between adjacent strings. In the following, these basic ingredients of the simulations – the computation of the random trajectories and the incorporation of the aggregation mechanisms – are described in detail.

**Simulation of the string trajectories.** In the absence of aggregation processes, the trajectory of the strings is determined by stochastic forces stemming from the HMM motor proteins at the surface. Similar to worm-like bundles (43), it can be expected that the curvature distribution is of approximately Gaussian shape. Mathematically such a distribution can be generated by a stochastic process defined by the following update rule for the curvature $\kappa$

$$\kappa(t_{n+1}) = 0.5 \cdot ( \kappa(t_n)+ \eta ), \tag{1}$$

whereby n denotes the time step and $\eta$ represents a random variable that is uniformly distributed over the interval $[-\alpha,\alpha]$. Stochastic changes of a given curvature occur with a rate $\omega$, with equally spaced time intervals, $t_{n+1} - t_n = \omega^{-1}$. The resulting distribution for the curvature is given in the supplemental figure S4.

**Modelling of the aggregation mechanisms.** Aggregation of the strings into longer and thicker bundles affects the strings' trajectories in a twofold way. While elongation of strings changes their probability to collide and form even longer strings or closed loops, thicker bundles that are the result of lateral aggregation, are less susceptible to curvature changes.

The elongation of strings is mainly based on merging events between adjacent strings, which occurs only for certain collision parameters: The colliding beads, say bead 1 and 2, have to be within a capture distance $R_1+R_2$, where $R_1$ and $R_2$ denote the radii of the particular beads. The bead's radius $R_i$ mainly depends on the excluded volume of the strings and scales with the number of already merged strings M according to $R_i = MR^0_i L_0/L$, where $R^0_i$ denotes the initial radius and $L$ the length of the structure. Further, strings only merge if the collision angles $|\theta|$ are smaller than a critical merging angle $\theta_c$. This accounts for the binding properties of fascin, which is only able to link approximately parallel orientated filaments (44). The merging naturally includes ring closure events, when the head of a string interacts with its own tail.

The lateral aggregation is based on the continuous uptake of individual filaments and small strings, leading predominantly to a growth in thickness. This results in an increased stiffness of the strings that is modelled by a linear increase of what we call a tenacity parameter $b(t)$

$$b(\tau_{n+1}) = b(\tau_n) + \beta b_0, \quad (2)$$

with $b_0$ being the initial tenacity $b(\tau=0)$ and $\beta$ determining the thickening speed. The time intervals are taken as equally spaced and $\tau_{n+1} - \tau_n = \lambda^{-1}$ defines the thickening rate $\lambda$. The linear tenacity increase modelled in Eq. (2) is motivated by a linear increase of the thickness of the strings. This implicitly implies an infinite filament reservoir. Since the experimentally observed ring formation is completed long before the filament reservoir is depleted, this is in good agreement with experimental observations. Thicker strings have a higher tenacity and are thus less susceptible to curvature changes. This can readily be modelled by modifying the update rule of Eq. (1) with a weight factor $w[b(t)]$ that depends on the tenacity

$$\kappa(t_{n+1}) = (w[b(t_n)] \cdot \kappa(t_n) + \eta) / (w[b(t_n)] + 1). \quad (3)$$

The functional dependence of the weight factor $w$ on the tenacity $b$ is given by

$$w[b(t)] = \exp([b(t)-b_0]/b_0). \quad (4)$$

**Parameters.** In all simulations, the initial bead radius was set to half of the unit length, and initially each string consisted of $N=10$ beads, implying and initial string length of $L_0=10$. The velocity was set to $v=1.0$, i.e. a string covers its own length in a simulation time of 10. All rates are measured in units of $v/L_0$. Thus a rate of 0.1 means that the corresponding stochastic process occurs on the time scale a string needs to cover its own length. For all simulations the thickening speed β was chosen to $\beta = 0.2$. For every simulation run, we started with 1500 strings distributed randomly in space and orientation in a cubic simulation box of size $L_{box}= 50 \cdot L_0$ with periodic boundary conditions. The time discretization $\Delta t$ is chosen to 0.1. The measured quantities are robust against changes in $\Delta t$ and the size of the simulation box. In all simulations, the total simulation time was chosen to $5 \cdot 10^4$ $\Delta t$, ensuring that all measured quantities are within a well saturated regime. The histograms and all values for $\Gamma$ are averaged over 20 simulation runs of randomized initial configurations and curvature amplitudes. For visualisation and image processing we used VMD and ImageJ.

**Acknowledgements**

We thank K. Voigt and M. Rusp for assistance in the protein purification. We gratefully acknowledge funding from the European Research Council under the European Union's Seventh Framework Programme (FP7/2007-2013)/ERC CompNet (279476). Financial support from the DFG in the framework of the SFB 863 and the German Excellence Initiatives via the 'Nano-Initiative Munich (NIM)' and the Technische Universität München - Institute for Advanced Study is gratefully acknowledged. V.S. and C.W. acknowledge support from the Elite Network of Bavaria by the graduate programmes CompInt and NanoBioTechnology and the International Graduate School of Science and Engineering (IGSSE).

**References**

1. Onsager L (1931) Reciprocal relations in irreversible processes. I *Phys Rev* 37:405.
2. Kudrolli A, Lumay G, Volfson D, Tsimring LS (2008) Swarming and swirling in self-propelled polar granular rods *Phys Rev Lett* 100:058001.
3. Narayan V, Ramaswamy S, Menon N (2007) Long-lived giant number fluctuations in a swarming granular nematic *Science* 317:105.
4. Ward AJW, Sumpter DJT, Couzin LD, Hart PJB, Krause J (2008) Quorum decision-making facilitates information transfer in fish shoals *Proc Natl Acad Sci USA* 105:6948.
5. Couzin ID, Krause J, Franks NR, Levin SA (2005) Effective leadership and decision-making in animal groups on the move *Nature* 433:513.
6. Ballerini M, Calbibbo N, Candeleir R, Cavagna A, Cisbani E, Giardina I, Lecomte V, Orlandi A, Parisi G, Procaccini A*, et al.* (2008) Interaction ruling animal collective behavior depends on topological rather than metric distance: Evidence from a field study *Proc Natl Acad Sci USA* 105:1232.
7. Cisneros LH, Cortez R, Dombrowski C, Goldstein RE, Kessler JO (2007) Fluid dynamics of self-propelled microorganisms, from individuals to concentrated populations *Exp Fluids* 43:737.
8. Dombrowski C, Cisneros L, Chatkaew S, Goldstein RE, Kessler JO (2004) Self-concentration and large-scale coherence in bacterial dynamics *Phys Rev Lett* 93:098103.
9. Riedel IH, Kruse K, Howard J (2005) A self-organized vortex array of hydrodynamically entrained sperm cells *Science* 309:300.
10. Nedelec FJ, Surrey T, Maggs AC, Leibler S (1997) Self-organization of microtubules and motors *Nature* 389:305.
11. Butt T, Mufti T, Humayun A, Rosenthal PB, Khan S, Molloy JE (2009) Myosin Motors Drive Long Range Alignment of Actin Filaments *J Biol Chem* 285:4964.
12. Schaller V, Weber C, Semmrich C, Frey E, Bausch AR (2010) Polar patterns of driven filaments *Nature* 467:73.
13. Köhler S, Schaller V, Bausch AR (2011) Structure formation in active networks *Nature Materials* 10:462.
14. Simha RA, Ramaswamy S (2002) Hydrodynamic fluctuations and instabilities in ordered suspensions of self-propelled particles *Phys Rev Lett* 89:058101.
15. Aranson IS, Tsimring LS (2005) Pattern formation of microtubules and motors: Inelastic interaction of polar rods *Phys Rev E* 71:050901.
16. Aranson IS, Sokolov A, Kessler JO, Goldstein RE (2007) Model for dynamical coherence in thin films of self-propelled microorganisms *Phys Rev E* 75:040901.
17. Baskaran A, Marchetti MC (2008) Hydrodynamics of self-propelled hard rods *Phys Rev E* 77:011920.
18. Baskaran A, Marchetti MC (2009) Statistical mechanics and hydrodynamics of bacterial suspensions *Proc Natl Acad Sci USA* 106:15567.
19. Cates ME, Fielding SM, Marenduzzo D, Orlandini E, Yeomans JM (2008) Shearing active gels close to the isotropic-nematic transition *Phys Rev Lett* 101:068102.
20. Liverpool TB, Marchetti MC (2006) Rheology of active filament solutions *Physical Review Letters* 97:268101.
21. Hatwalne Y, Ramaswamy S, Rao M, Simha RA (2004) Rheology of active-particle suspensions *Phys Rev Lett* 92:118101.
22. Angelini TE, Hannezo E, Trepat X, Marquez M, Fredberg JJ, Weitz DA (2011) Glass-like dynamics of collective cell migration *Proc Natl Acad Sci USA* 108:4714.
23. Umbanhowar PB, Melo F, Swinney HL (1996) Localized excitations in a vertically vibrated granular layer *Nature* 382:793.
24. Olafsen JS, Urbach JS (1998) Clustering, Order, and Collapse in a Driven Granular Monolayer *Phys Rev Lett* 81:4369.

25. Aranson IS, Tsimring LS (2006) Patterns and collective behavior in granular media: Theoretical concepts *Rev Mod Phys* 78:641.
26. Tsai JC, Ye F, Rodriguez J, Gollub JP, Lubensky TC (2005) A Chiral Granular Gas *Phys Rev Lett* 94:214301.
27. Hinrichsen H (2000) Non-equilibrium critical phenomena and phase transitions into absorbing states *Adv Phys* 49:815.
28. Henkel M, Hinrichsen H, Lubeck S (2009) *Non-Equilibrium Phase Transitions: Volume 1: Absorbing Phase Transitions* (Springer Netherlands).
29. Hallatschek O, Nelson DR (2009) Life at the front of an expanding population *Evolution* 64:193.
30. Corte L, Chaikin PM, Gollub JP, Pine DJ (2008) Random organization in periodically driven systems *Nature Physics* 4:420.
31. Hess H, Clemmens J, Brunner C, Doot R, Luna S, Ernst KH, Vogel V (2005) Molecular self-assembly of "nanowires" and "nanospools" using active transport *Nano Letters* 5:629.
32. Liu HQ, Spoerke ED, Bachand M, Koch SJ, Bunker BC, Bachand GD (2008) Biomolecular Motor-Powered Self-Assembly of Dissipative Nanocomposite Rings *Advanced Materials* 20:4476.
33. Takeuchi KA, Kuroda M, Chat, eacute, Hugues, Sano M (2007) Directed Percolation Criticality in Turbulent Liquid Crystals *Physical Review Letters* 99:234503.
34. Liu C-H, Pine DJ (1996) Shear-Induced Gelation and Fracture in Micellar Solutions *Physical Review Letters* 77:2121.
35. Schmoller KM, Fernandez P, Arevalo RC, Blair DL, Bausch AR (2010) Cyclic hardening in bundled actin networks *Nature Communications* 1:134.
36. Spudich JA, Watt S (1971) Regulation of rabbit skeletal muscle contraction. 1. Biochemical studies of interaction of Tropomyosin-Troponin complex with Actin and proteolytic fragments of Myosin *J Biol Chem* 246:4866.
37. MacLean-Fletcher S, Pollard TD (1980) Identification of a factor in conventional muscle Actin preparations which inhibits Actin filament self-association *Biochem. Biophys Res Commun* 96:18.
38. Margossian SS, Lowey S (1982) Preparation of Myosin and its subfragments from rabbit skeletal muscle *Methods Enzymol* 85:55.
39. Vignjevic D, Yarar D, Welch MD, Peloquin J, Svitkina T, Borisy GG (2003) Formation of filopodia-like bundles in vitro from a dendritic network *J Cell Biol* 160:951.
40. Toyoshima YY, Kron SJ, Spudich JA (1990) The Myosin step-size - measurement of the unit displacement per ATP hydrolyzed in an invitro assay *Proc Natl Acad Sci USA* 87:7130.
41. Sundberg M, Balaz M, Bunk R, Rosengren-Holmberg JP, Montelius L, Nicholls IA, Omling P, Tagerud S, Mansson A (2006) Selective spatial localization of actomyosin motor function by chemical surface patterning *Langmuir* 22:7302.
42. Schilling J, Sackmann E, Bausch AR (2004) Digital imaging processing for biophysical applications *RevSci Instrum* 75:2822.
43. Heussinger C, Bathe M, Frey E (2007) Statistical mechanics of semiflexible bundles of wormlike polymer chains *Phys RevLett* 99:048101.
44. Courson DS, Rock RS (2010) Actin Cross-link Assembly and Disassembly Mechanics for alpha-Actinin and Fascin *J Biol Chem* 285:26350.

**Figures**

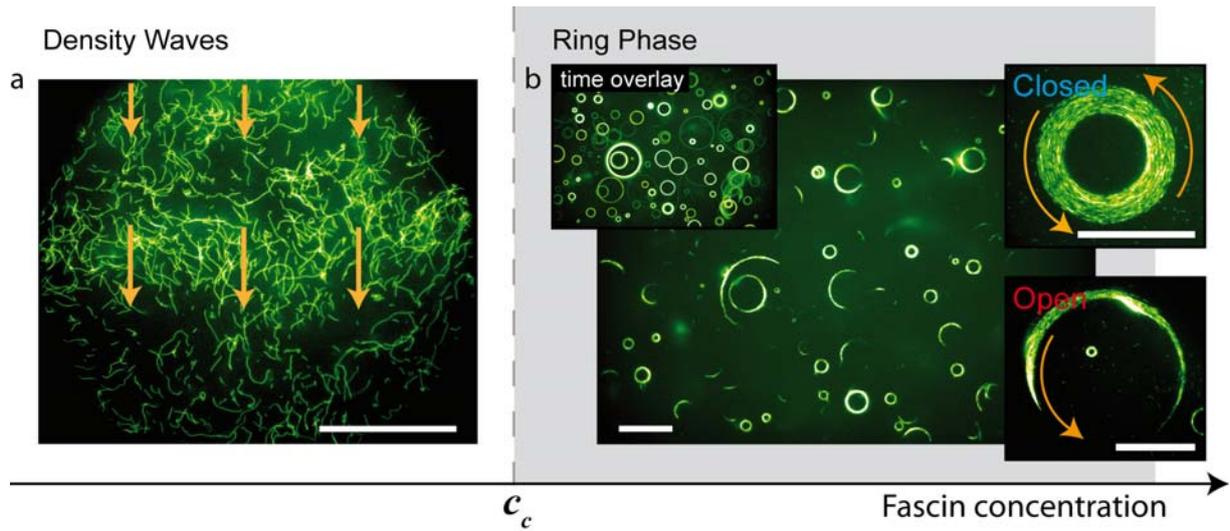

**Figure 1. Phase behaviour as a function of the fascin concentration.** Low fascin concentrations are not sufficient to fundamentally alter the pattern formation in the high density motility assay and at an actin concentration of $\rho = 10\mu M$ the characteristic travelling density waves evolve (*A*). Compared to the case without crosslinkers they are less pronounced as the crosslinker slightly hinders the formation of density inhomogeneities. Above a critical fascin concentration of $c_c = 0.075 \pm 0.025$ μM the pattern formation drastically changes as constantly rotating rings evolve (*B*) that are either closed or open (supplemental movie S1). In the steady state all actin filaments are incorporated in rotating rings and fluctuations on the single filament level are entirely absent. All scale bars are 50μm.

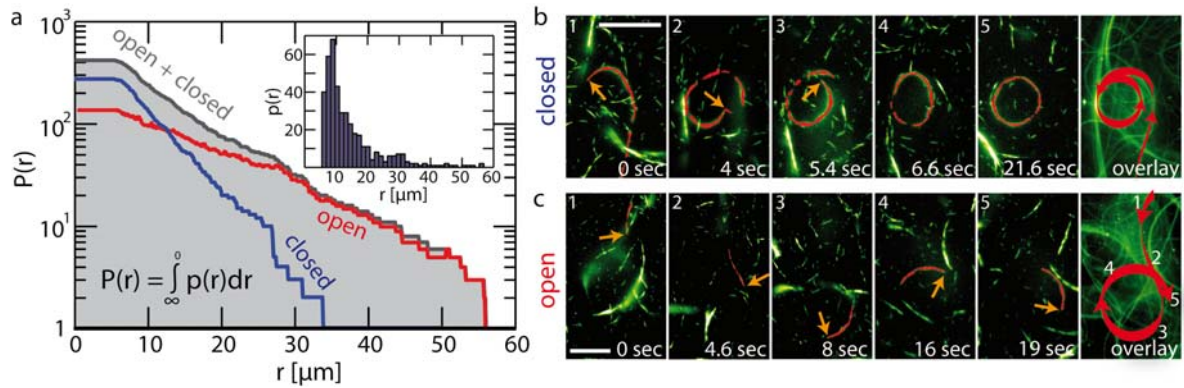

**Figure 2. Ring curvature distributions and ring formation mechanisms.** (*A*) shows the cumulative curvature radii distribution $P(r)$ in the frozen steady state; the inset depicts the non-cumulative distribution $p(r)$. The distribution can be described by a double exponential decay according to $P(r) \propto A_1 \exp(r/l_1) + A_2 \exp(r/l_2)$ with decay lengths of $l_1 = 3.3$ μm and $l_2 = 10.1$ μm. This double exponential shape reflects the occurrence of two different ring morphologies: open and closed rings. These two different ring populations rely on distinct ring formation mechanisms that are related to the growth mechanisms in the system. While being transported, moving actin-fascin strings grow by merging with other strings of similar size. This leads to elongated but still flexible strings that predominantly form closed ring if they cross their own tail (*B*). Open rings form upon a different mechanism (*C*): While moving, actin fascin strings continuously pick up material – individual filaments or smaller actin fascin strings. Thereby they grow predominately in width and get thicker and stiffer. If they are stiff enough the curvature freezes and the forces and fluctuations in the motility assay are not sufficient anymore to induce any change in curvature. While closed rings characteristically are small in size with radii of up to 30 μm, open rings are considerably broader and can have radii of up to 150 μm (*A*). The actin concentration was set to $\rho = 3$ μM and the fascin concentration was $c = 0.2$ μM. All scale bars are 50μm. In (*B*) and (*C*) the investigated actin-fascin string is shown in red and its tip is marked by a yellow arrow.

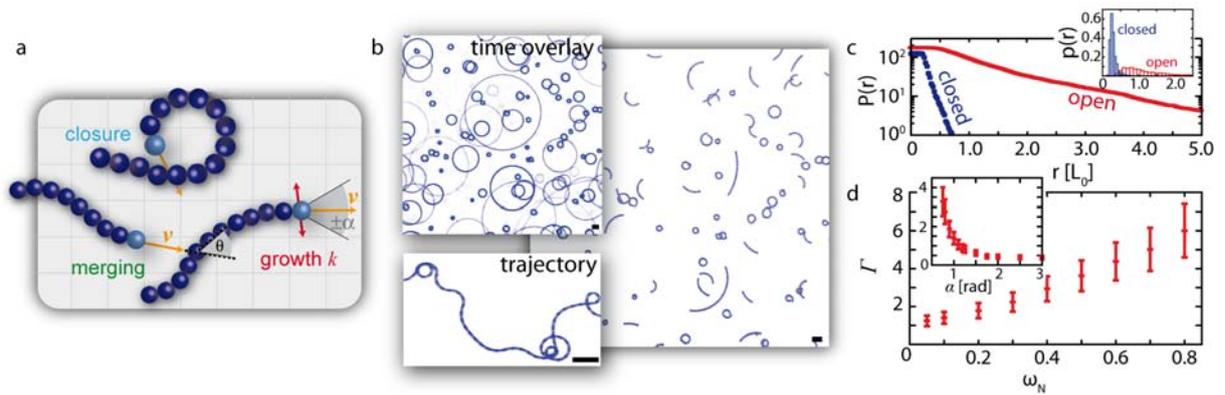

**Figure 3. Cellular automaton simulations.** The two experimentally observed aggregation processes – merging and growth stiffening are described using a continuous agent based simulation. The actin fascin strings are modelled as polar elongated strings that move with with a velocity *v* on meandering trails (*A*). The tip is subjected to curvature changes of rate $\omega$ and a noise level $\alpha$, resulting in a meandering trajectory (inset *B*). The strings stiffen due to growth processes of rate $\lambda$. Besides, merging events with adjacent objects occur if their relative angle $|\theta|$ is smaller than $\theta_c$. These two aggregation processes lead to the emergence of rings in two configurations – open and closed (*B*). This is reflected in the ring radii distribution *p(r)* and the corresponding cumulative distribution *P(r)* that can be separated in open and closed contributions (*C*). The distribution for open and closed rings decays approximately exponential. The ratio of open to closed rings $\Gamma$ increases with the rate of the random turns $\omega$, while it decreases with the noise level $\alpha$ (*D*). If not indicated otherwise the parameters are $\omega = 0.1$, $\lambda = 0.4$, $\alpha = 1.0$ and $\theta_c = 10^0$. All scale bars are one string length $L_0$.

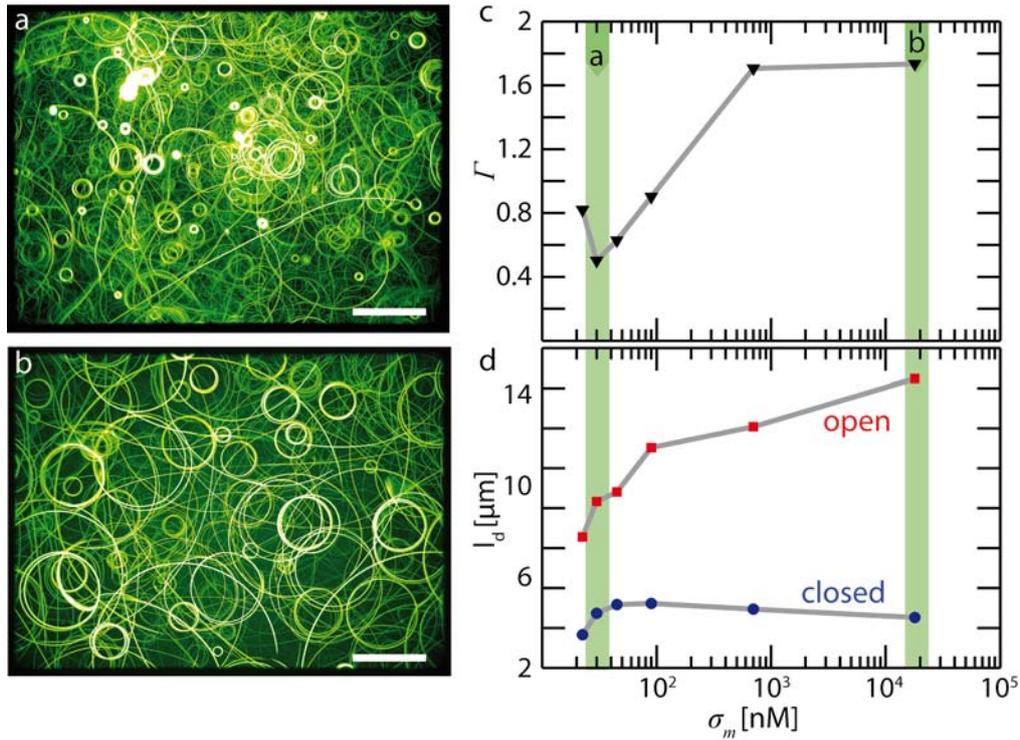

**Figure 4. Dependence on the motor density.** By varying the motor density $\sigma_m$ on the surface, the feedback between growth, and the fluctuations that arise in the motility assay can be examined. In general high motor densities lead to less fluctuations and a more persistent movement as can be seen in the time overlay image in (*B*), $\sigma_m$ = 700 nM, while low motor densities lead to a more fluctuations (*A*), $\sigma_m$ = 30 nM. The higher the noise level (i.e. the lower $\sigma_m$) the higher is the chance for individual strings to cross their own tail and to form closed rings. This is directly reflected by the higher abundance of closed rings at low motor densities quantified by the ratio of open to closed rings $\Gamma$ (*C*). The radial distribution of closed rings itself is unaffected as can be seen in the decay lengths $l_d$ of the exponentially decaying distributions (*D*). This is different for open rings, where the frozen-in curvature of the open rings directly reflects the fluctuations and the persistence of the movement in the motility assay: here $l_d$ monotonically increases with increasing motor concentration $\sigma_m$. The actin concentration was set to $\rho$ = 3 µM and the fascin concentration was $c$ = 0.2 µM. All scale bars are 50µm.

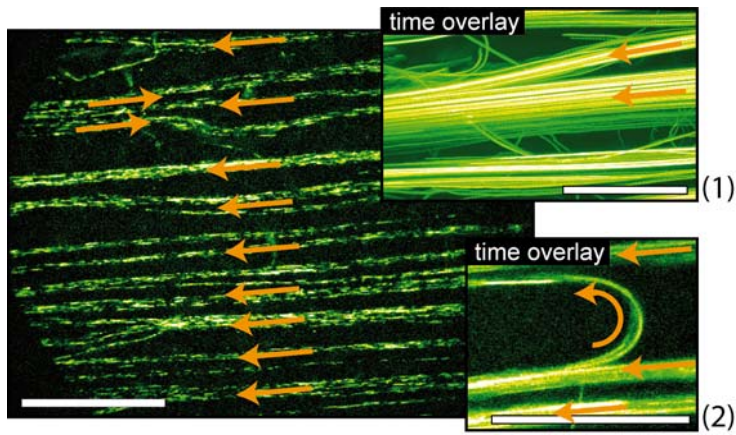

**Figure 5. Collectively moving streaks.** Above a certain material density, the ring formation processes are hindered and elongated actin fascin streaks emerge that are aligned alongside the long axis of the flow chamber. Like in the ring phase, all individual filaments gradually get incorporated in large scale polar structures and fluctuations on the single filament level cease. Initially the streaks move contrariwise. With time, a major direction of motion develops that is adopted by all streaks in the course of turning events (inset 2). This is accompanied by a coarsening process, where polarly aligned fibres gradually merge to larger ones (inset 1). The actin concentration was set to $\rho = 10\mu M$ and the fascin concentration was $c = 0.5\mu M$. All scale bars are 50µm.

# Frozen steady states in active systems:
# Supporting Material

Volker Schaller, Benjamin Hammerich and Andreas R. Bausch
*Lehrstuhl für Biophysik-E27, Technische Universität München, Garching, Germany*

Christoph Weber and Erwin Frey
*Arnold Sommerfeld Center for Theoretical Physics and CeNS, Department of Physics, Ludwig-Maximilians-Universität, Munich, Germany*

05.09.2011

## Supplemental Figures

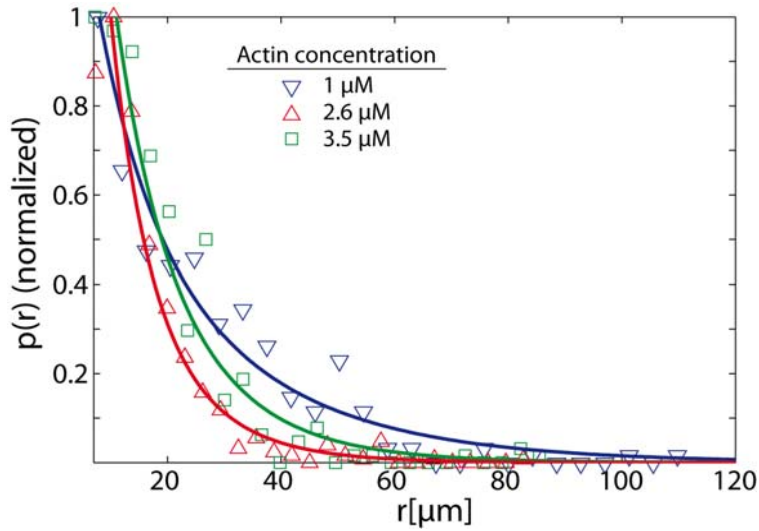

**Supplemental Figure S1:** Decay of the curvature radii as a function of the actin concentration. Independent of the actin concentrations the distributions decay according to $p(r) = a_1 \cdot \exp(-r/l_1) + a_2 \cdot \exp(-r/l_2)$, shown as solid lines. For the lowest actin concentration (1 μM, blue curve) parameters are $a_1 = 0.59$; $a_2 = 0.41$; $l_1 = 9.34$; $l_2 = 24.62$. For the intermediate actin concentration (2.6 μM, red curve) $a_1 = 0.53$; $a_2 = 0.47$; $l_1 = 5.10$; $l_2 = 12.57$ was found and for the highest actin concentration (3.5 μM, green curve) $a_1 = 0.59$; $a_2 = 0.41$; $l_1 = 8.29$; $l_2 = 15.01$ resulted. The motor density was adjusted to $\sigma_m = 90$ nM and the fascin concentration was $c = 0.5$ μM.

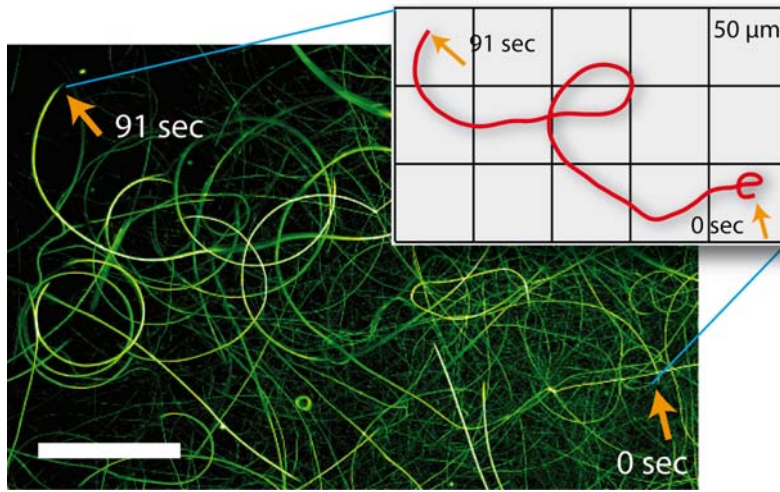

**Supplemental Figure S2:** Time overlay of a moving actin-fascin string. Actin-fascin strings move on circular trajectories. The thicker the strings get, the higher their directional persistence. As a consequence variations of a given curvature happen increasingly less frequent and have less effect. The motor density was adjusted to $\sigma_m$ = 90 nM, the actin concentration was $\rho$ = 3 µM and the fascin concentration was $c$ = 0.5 µM. The scalebar is 50 µm.

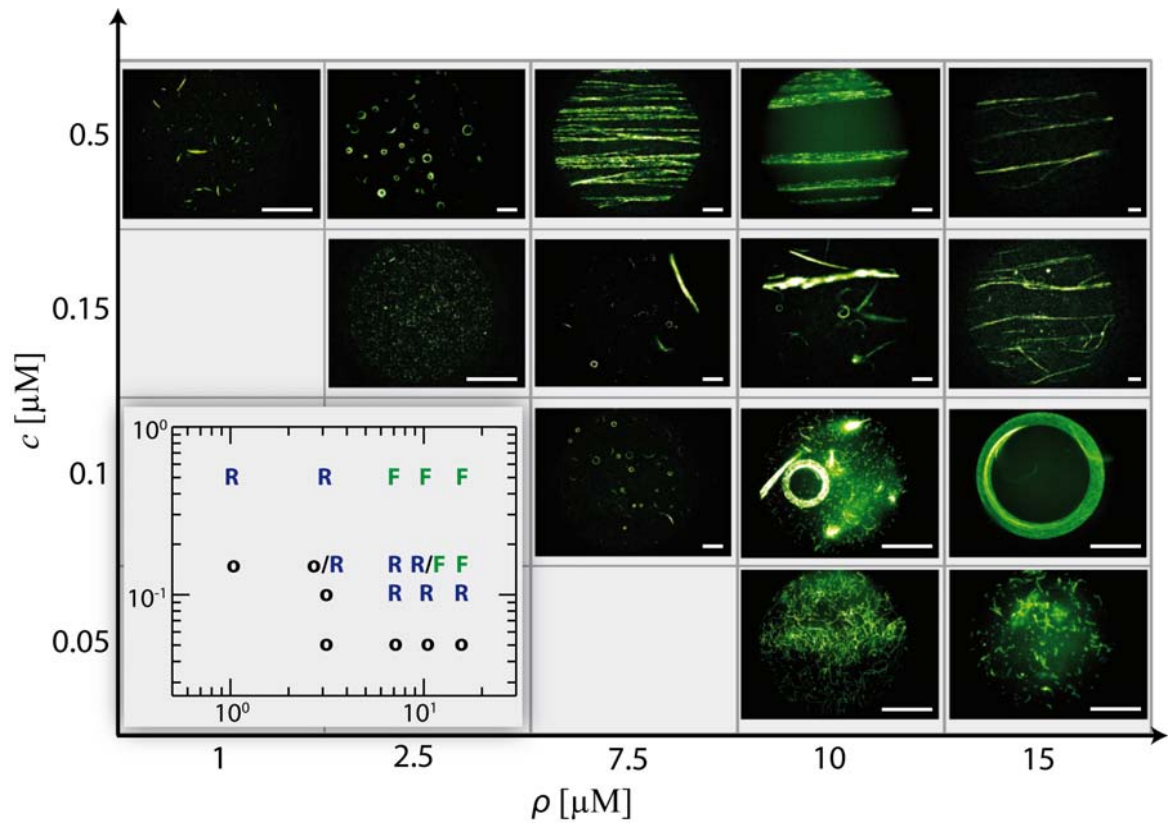

**Supplemental Figure S3:** Phase diagram as a function of the actin $\rho$ and fascin concentration $c$. For high actin and fascin concentration the systems evolves into a frozen steady state that is characterized by coherently moving streaks or fibres (**F**). For intermediate concentrations of actin and fascin the frozen active state is given by the ring phase (**R**). At low fascin concentrations the steady state is characterized by persistent fluctuations on the single filament level and no frozen steady state emerges (**o**). All scale bars are 50 μm and the motor density was adjusted to 90 nM.

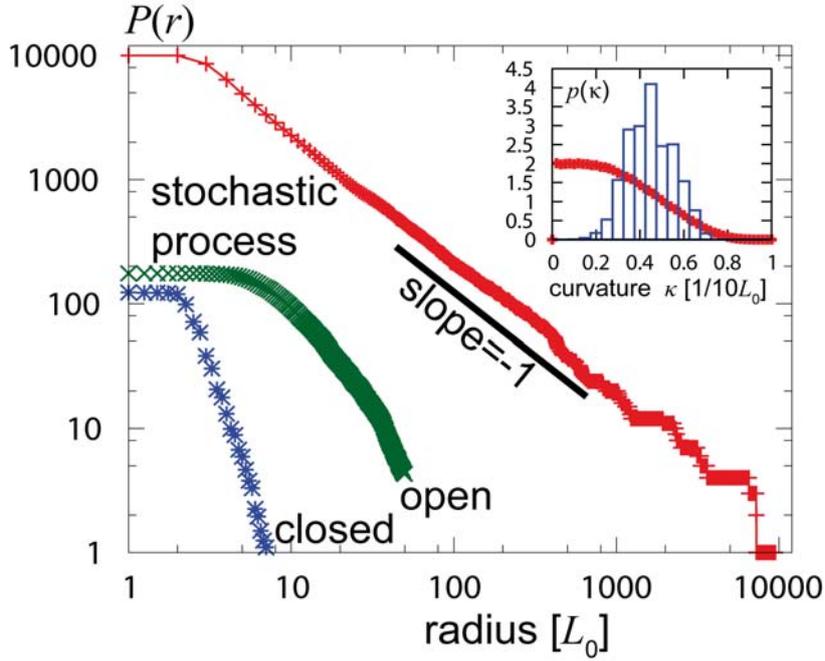

**Supplemental Figure S4.** Cumulative radii distribution P(r) with and without aggregation. The stochastic process determined by equation (1) in the main manuscript leads to a power-law of slope -1 (black line) in the cumulative radii distribution P(r) (red curve). In contrast, simulations that include growth stiffening according to equations (3) and (4), as well as the possibility of merging events, exhibit a different shape (green and blue curve). Inset: Normalized non-cumulative curvature distribution $p(\kappa)$ for stochastic processes according to equation (1) (red curve) and for the aggregation process according to equation (2) (blue boxes correspond to closed). In all simulation runs $\alpha$ was set to $\alpha =1.0$. The parameters that determine the aggregation processes were set to: $\lambda =0.1$, $\omega =0.4$, $\theta_c =10^0$ (see blue and green curve).

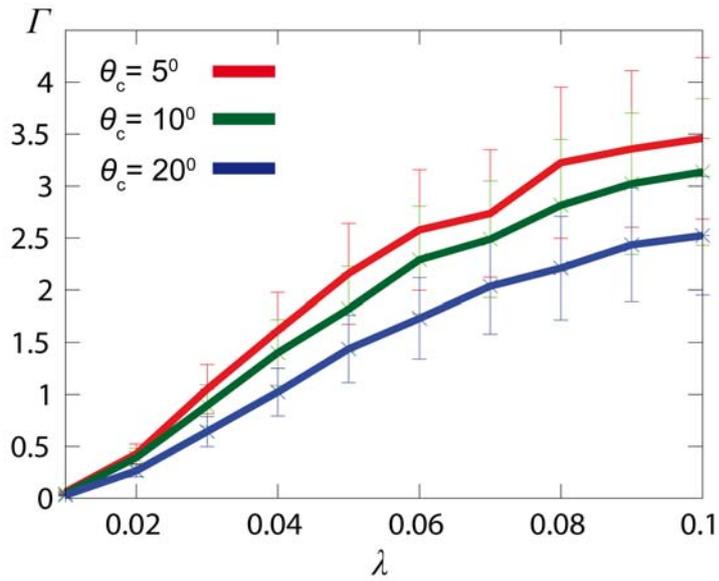

**Supplemental Figure S5**. Dependence of $\Gamma$ on growth rate $\lambda$ for three critical merging angles $\theta_c$. The ratio of open to closed rings $\Gamma$ increases with the thickening rate $\lambda$. Slow growth stiffening allows the emergence of longer structures leading to more closed ring, whereas fast growth stiffening freezes the structure's orbits resulting in open rings spatially separated without any further interactions. The critical merging angle has only a minor impact on the ring statistics. The parameters characterizing the noise level are: $\omega = 0.1$, $\alpha = 1.0$.

## Supplemental Movies: Movie Captions

**Supplemental Movie S1:** The addition of fascin leads to the emergence of a frozen steady state of rotating rings (actin concentration $\rho = 3$ µM, motor concentration $\sigma_m = 90$ nM, fascin concentration $c = 0.2$ µM, labeling ratio 1:16).

**Supplemental Movie S2:** The emergence of stable curvatures relies on two mechanisms: the ring closure and the freezing of a current curvature (actin concentration $\rho = 3$ µM, motor concentration $\sigma_m = 90$ nM, fascin concentration $c = 0.2$ µM, labeling ratio 1:16).

**Supplemental Movie S3:** Emergence of rotating rings in the cellular automaton simulations and coexistence of open and closed rings ($\omega = 0.1$, $\lambda = 0.4$, $\alpha = 1.0$ and $\theta_c = 10^0$).

**Supplemental Movie S4:** Above a certain material density curved trajectories are no longer possible and the actin fascin structures are forced on straight trajectories; extended actin-fascin streaks evolve that successively merge and coarsen, leading to a large scale symmetry breaking (actin concentration $\rho = 10$ µM, motor concentration $\sigma_m = 90$ nM, fascin concentration $c = 0.5$ µM, labeling ratio 1:32).